\documentclass[aps,prl,reprint,amsmath,amssymb,a4paper,floatfix,showpacs,citeautoscript,superscriptaddress]{revtex4-1}
\usepackage{mathptmx}
\usepackage[utf8]{inputenc} 
\usepackage{subfigure}
\usepackage{graphicx}
\usepackage{epstopdf}
\usepackage{xspace}
\usepackage[
,textwidth=17.5cm
,textheight=23.0cm 
,verbose
,dvips
]{geometry}

\usepackage{color}

\begin{document}

\title{Computing Equilibrium Shapes of Wurtzite Crystals: The Example of GaN}

\author{Hong Li}
\email{hong.li@physik.hu-berlin.de}
\affiliation{Institut f\"ur Physik and IRIS Adlershof, Humboldt-Universit\"at zu Berlin, Zum Gro\ss en Windkanal 6, 12489 Berlin, Germany}
\affiliation{Paul-Drude-Institut f\"ur Festk\"orperelektronik, Hausvogteiplatz 5-7, 10117 Berlin, Germany}
\author{Lutz Geelhaar}
\affiliation{Paul-Drude-Institut f\"ur Festk\"orperelektronik, Hausvogteiplatz 5-7, 10117 Berlin, Germany}
\author{Henning Riechert}
\affiliation{Paul-Drude-Institut f\"ur Festk\"orperelektronik, Hausvogteiplatz 5-7, 10117 Berlin, Germany}
\author{Claudia Draxl}
\affiliation{Institut f\"ur Physik and IRIS Adlershof, Humboldt-Universit\"at zu Berlin, Zum Gro\ss en Windkanal 6, 12489 Berlin, Germany}

\begin{abstract} 
Crystal morphologies are important for the design and functionality of devices based on low-dimensional nanomaterials. The equilibrium crystal shape (ECS) is a key quantity in this context. It is determined by surface energies, which are hard to access experimentally but can generally be well predicted by first-principles methods. Unfortunately, this is not necessarily so for polar and semipolar surfaces of wurtzite crystals. By extending the concept of Wulff construction, we demonstrate that ECSs can nevertheless be obtained for this class of materials. For the example of GaN, we identify different crystal shapes depending on the chemical potential, shedding light on experimentally observed GaN nanostructures.
\end{abstract}


\maketitle
Low-dimensional semiconductor nanostructures have attracted a lot of interest in the past decades, largely due to their  applications in  low-energy consumption and energy-harvesting devices \cite{Law2004,Pearton2008}. Owing to surface effects, the performance of such devices strongly depends on the nanocrystal morphology. To achieve comprehensive understanding and control of the preferred growth morphology, one must know the material's natural shape that results from its crystallographic anisotropy. {\it Ab initio} theory can generally provide more insight into this complicated issue through the calculation of surface energies since, according to Wulff's theorem \cite{Wulff1901}, the equilibrium crystal shape (ECS) of a solid can be constructed by the mere knowledge of surface energies of various crystal planes. For a crystalline solid, the surface energy $\gamma$ is defined as the excess free energy required to create one unit of surface area $A$ \cite{Gibbs1957},
\begin{equation}
\gamma=\frac{1}{A}[G-\sum_i N_i \mu_i].
\label{eq:gamma}
\end{equation}
$G$ represents the Gibbs free energy of the system that, neglecting temperature and pressure, is replaced by the total energy. The chemical potential $\mu_i$ is the free energy per atom in the system for species \textit{i}, and $N_i$ denotes the number of atoms of this species. In a bulk material, the total chemical potential is known from the corresponding total energy $E_\text{bulk}=\sum_in_i\mu_i$, where $n_i$ is the number of atoms of species $i$ in the bulk. Hence, the surface energy of a nonpolar plane can be extracted from density-functional-theory (DFT) results for a slab that contains two identical surfaces well separated from each other. For some polar and semipolar planes, however, individual surface energies are difficult to access, because different facets may appear at the two surfaces of the slab. To overcome this problem, a method has been proposed \cite{Zhang2004} involving two surface types on three side faces of a triangular wedge. This approach is, however, not applicable to all surfaces and crystal structures; polar surfaces in wurtzite crystals are one example \cite{Ruterana2003,footnote}.  Consequently, not every individual surface energy of wurtzite crystals can be computed; hence, the construction of the ECS seemed impossible. Recently, neglecting the different layer-stacking sequence, the surface energy of the polar \textit{c} plane was estimated from the zincblende (111) plane \cite{Dreyer2014}.

In this Letter, we show that such an approximation is not required to unambiguously determine the ECS. We introduce a generalization of the Wulff construction, based on combinations of surface energies, to show how the ECS for the class of wurtzite materials can be obtained. We demonstrate this principle by taking GaN as a technologically important example. The wide-band-gap semiconductor GaN is a key material in today's white-light-emitting diodes for general illumination, blue lasers, and high-power and high-frequency electronics \cite{Service2010}. GaN readily grows in the form of nanowires (NWs) in molecular beam epitaxy (MBE) \cite{Geelhaar2011,Consonni2013} and metal-organic chemical vapor deposition (MOCVD) \cite{Hersee2006,Chen2010}. However, different shapes are observed, depending on the growth temperature, pressure, and chemical environment \cite{Du2005,Jindal2009,Bryant2013,Urban2013,Jin2013}. Our study leads to new understanding of these GaN crystal shapes under various growth conditions. 

We performed DFT calculations in the local-density approximation using the projector-augmented-wave method \cite{Blochl1994} as implemented in  VASP code \cite{Kresse1996,Kresse1999}. (For computational details see Supplemental Material (SM) \cite{supp}.) The crystal planes considered here were chosen on the basis of experimental data and include the nonpolar \textit{m} plane $(1\overline{1}00)$ and \textit{a} plane $(11\overline{2}0)$, the polar \textit{c} planes (0001) and $(000\overline{1})$, and the semipolar planes $(11\overline{2}2)$, $(11\overline{22})$, $(1\overline{1}01)$, $(1\overline{1}0\overline{1})$, $(1\overline{1}02)$, and $(1\overline{1}0\overline{2})$. Six differently oriented slabs are constructed to calculate the surface energies according to Eq.~\eqref{eq:gamma}. Since the surface can have different terminations, below we label a surface by its plane indices together with a subscript of the terminating layer or bilayer. For the $a$ plane  and $m$ plane, individual surface energies are directly determined, because the two surfaces of the slab are identical. For the other slabs only the sum (average) of the surface energies of the two sides can be obtained. Depending on the two surface terminations, these slabs can be stoichiometric or nonstoichiometric, where, for the former, the sum of surface energies is independent of the chemical potential. Having considered two surface terminations for either side, we now determine the energetically most favorable combinations. For the [0001] slab,  $\gamma_\text{av}$ of $(0001)_\text{Ga}$ and $(000\overline{1})_\text{N}$ is lower than that of $(0001)_\text{N}$ and $(000\overline{1})_\text{Ga}$. For the $[11\overline{2}2]$ slab, the combination of $(11\overline{2}2)_\text{Ga}$ and $(11\overline{22})_\text{N}$ is more stable than that of  $(11\overline{2}2)_\text{N}$ and $(11\overline{22})_\text{Ga}$. Further, we identify $(1\overline{1}01)_\text{2N}$ and $(1\overline{1}0\overline{1})_\text{2Ga}$ to be favorable over $(1\overline{1}01)_\text{2Ga}$ and $(1\overline{1}0\overline{1})_\text{2N}$. Surface terminations of $(1\overline{1}02)_\text{GaN}$ and $(1\overline{1}0\overline{2})_\text{GaN}$ exhibit lower $\gamma_\text{av}$ than $(1\overline{1}02)_\text{Ga}$ combined with $(1\overline{1}0\overline{2})_\text{N}$.  The respective minimum average surface energies are summarized in Table~\ref{tab:SE} together with the individual values for the $a$ and $m$ planes. In agreement with previous DFT results \cite{Northrup1996}, we find the \textit{m} plane more stable than the \textit{a} plane by 8 meV/\AA$^2$. The average surface energies of the polar and semipolar planes are higher than those of nonpolar planes, indicating that the preferential growth is mainly along the \textit{c} axis, as always observed for GaN nanowires.

\begin{table}
\caption{Average surface energies  (in meV/\AA$^2$) of two opposite surfaces (relaxed) obtained from differently oriented slabs. The corresponding surface terminations are described in the text.}
\centering
\begin{tabular*}{\linewidth}{@{\extracolsep{\fill}} c  c  c  c  c  c  c }
\hline \hline
 Slab & $[1\overline{1}00]$ & $[11\overline{2}0]$ & $[0001]$ & $[11\overline{2}2]$ & $[1\overline{1}01]$ & $[1\overline{1}02]$ \\
$\gamma_\text{av}$ & 124 & 132 & 209 & 223 & 249 & 208 \\
\hline \hline
\end{tabular*}
\label{tab:SE}
\end{table}

To construct the ECS, we need surface energies to solve the equation
\begin{equation}
\begin{split}
r(\textbf{h})=\text{min}_{\textbf{m}}\left(\frac{\gamma(\textbf{m})}{\textbf{m}\cdot\textbf{h}}\right),
\end{split}
\label{eq:wc}
\end{equation}
where $r(\textbf{h})$ denotes the radius of the crystal shape along a given vector $\textbf{h}$ and $\gamma(\textbf{m})$ denotes the surface energy of a plane with normal vector $\textbf{m}$. Because individual surface energies are not accessible for wurtzite crystals, a straightforward solution of Eq.~\eqref{eq:wc} is not possible. However, an alternative geometrical route for determining the ECS can be accomplished using suitable surface-energy combinations. We will show below these quantities can be calculated. We note at this point that in any Wulff construction based on DFT, the different surfaces contributing to the ECS are obtained from individual calculations, i.e., charge redistribution between these surfaces towards a common Fermi level \cite{Ashcroft1976} is neglected. Possible effects related to Fermi-level pinning by surface sates, surface electrostatics, and reconstruction are briefly described in the SM \cite{supp}, and will be discussed elsewhere \cite{new_paper}. Estimates on the effect of surface reconstructions indicate that the shapes are hardly affected \cite{shapes}.

\begin{figure}
\includegraphics[width=\linewidth]{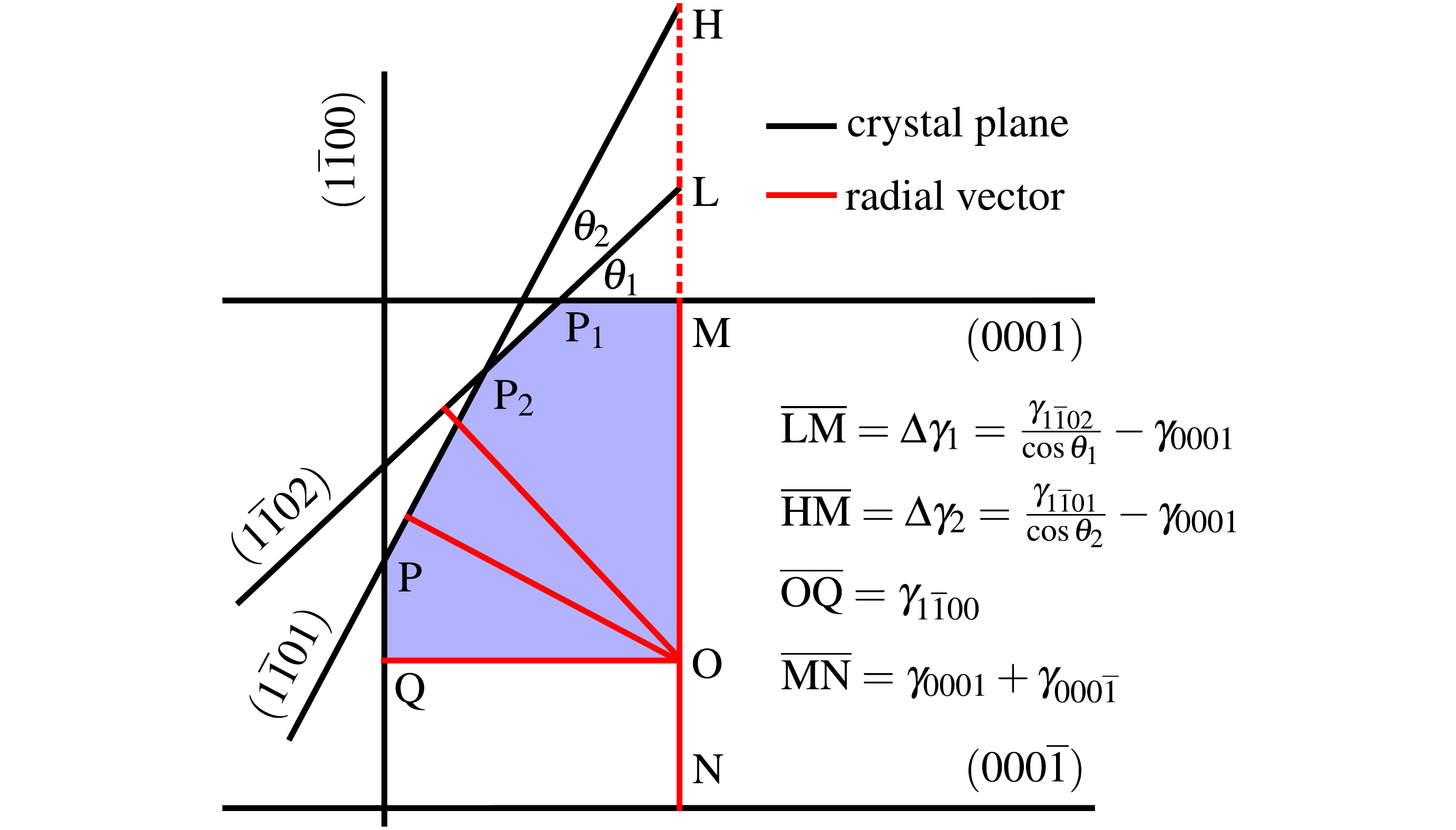}
\caption{\label{fig:Wulff} (color online) Schematic of a 2D Wulff construction. The crystal planes and radial vectors are depicted by black and red solid lines, respectively. The shaded area indicates the resulting quarter of the ECS.}
\end{figure}

To illustrate our central idea, Fig.~\ref{fig:Wulff} depicts a two-dimensional (2D) schematic for a generalized Wulff construction. With $\gamma_{0001} + \gamma_{000\overline{1}}$, the distance $\overline{MN}$ between these two planes is fixed, but their position with respect to the origin $O$ is not. This uncertainty, however, does not influence the crystal shape, because the surface-energy combinations actually fix its inner envelope. For instance, $\Delta \gamma_1=\gamma_{1\overline{1}02}/\cos \theta_1-\gamma_{0001}$, corresponding to the distance $\overline{LM}$, provides the difference in {\it hypothetical} crystal radii of the $(1\overline{1}02)$ plane, $\overline{OL}$, and the $(0001)$ plane, $\overline{OM}$, along the [0001] direction. ("Hypothetical" refers to the fact that the individual surface energies are not known.) The crossing point $P_1$ of these two planes is determined by $\overline{LM}$ and the dihedral angle $\theta_1$ (given by the lattice parameter). Likewise, $\Delta \gamma_2=\gamma_{1\overline{1}01}/\cos \theta_2-\gamma_{0001}$ fixes the length $\overline{HM}$ and, thus, together with $\theta_2$  we know the crossing point $P_2$. Since $\overline{OQ}$ can be calculated directly, the crossing point $P$ is clear. Overall, one can determine a quarter of the crystal shape, i.e., the shaded area in Fig. \ref{fig:Wulff}, despite the vertical coordinates of these points being unknown. The lower left quarter is constructed analogously, while the right half is determined by symmetry. Adding a constant value to an individual surface energy will reduce its counterpart by the same amount, resulting in a shift of the entire ECS along the \textit{c} axis. In fact, it was shown already in 1975 \cite{Arbel1975} for ten point groups that, for this reason, the determination of the ECS is not prevented. However, until the present, no way of constructing an ECS for such cases has been demonstrated.

\begin{figure}
\includegraphics[width=\linewidth]{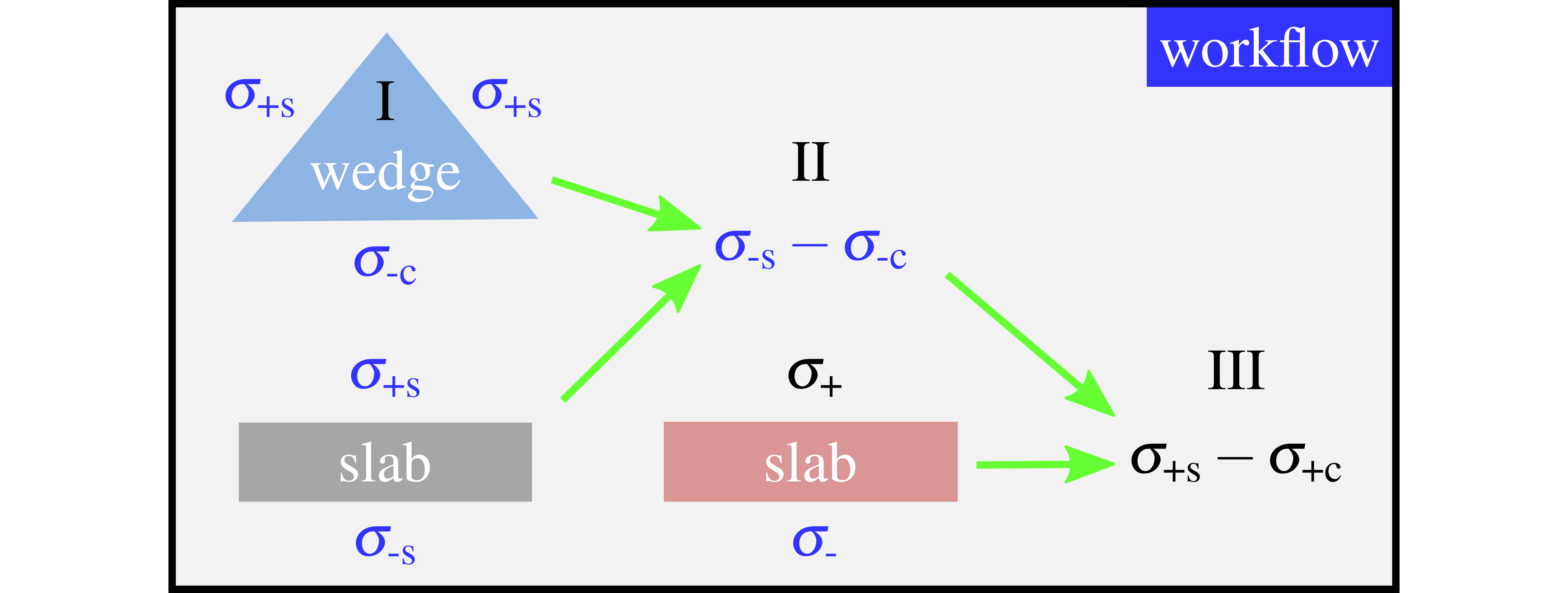}
\caption{(color online) Workflow of surface-energy calculations. Slabs and wedges are depicted by rectangular and triangular boxes, respectively. The surface energy $\sigma$ is labeled for each facet. The subscripts "s" and "c" stand for semipolar facet and polar facet, respectively, while "+" and "-" distinguish the facet orientation with respect to the [0001] direction. Surface-energy combinations are expressed symbolically, omitting the respective plane indices. \label{fig:workflow}}
\end{figure}

\begin{figure}
\subfigure{\label{fig:sa}\includegraphics[width=\linewidth]{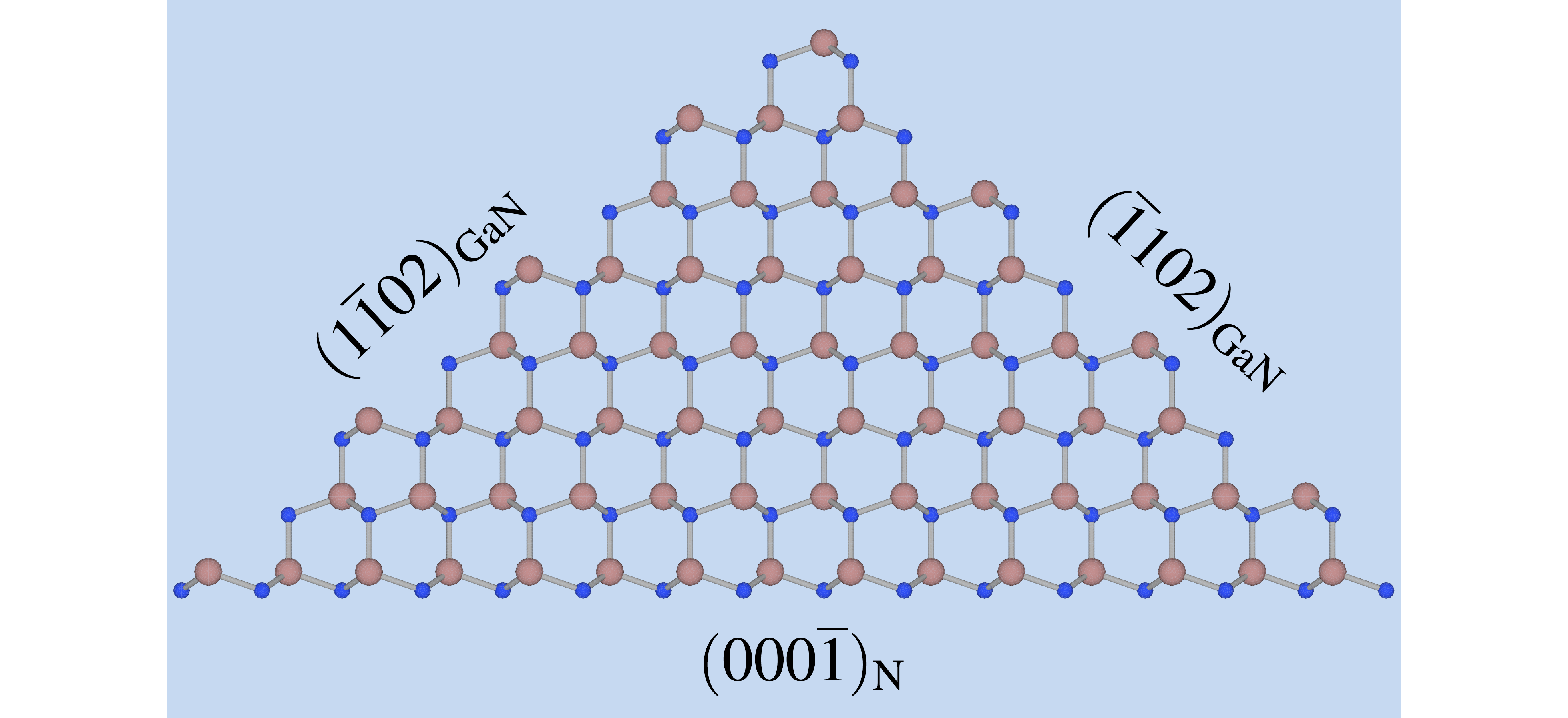}}
\subfigure{\label{fig:sb}\includegraphics[width=\linewidth]{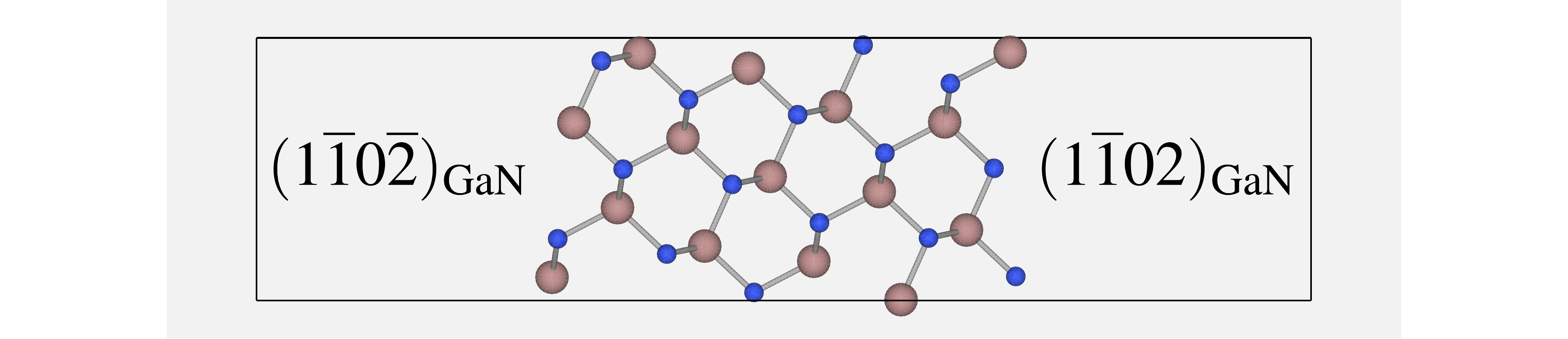}}
\subfigure{\label{fig:sc}\includegraphics[width=\linewidth]{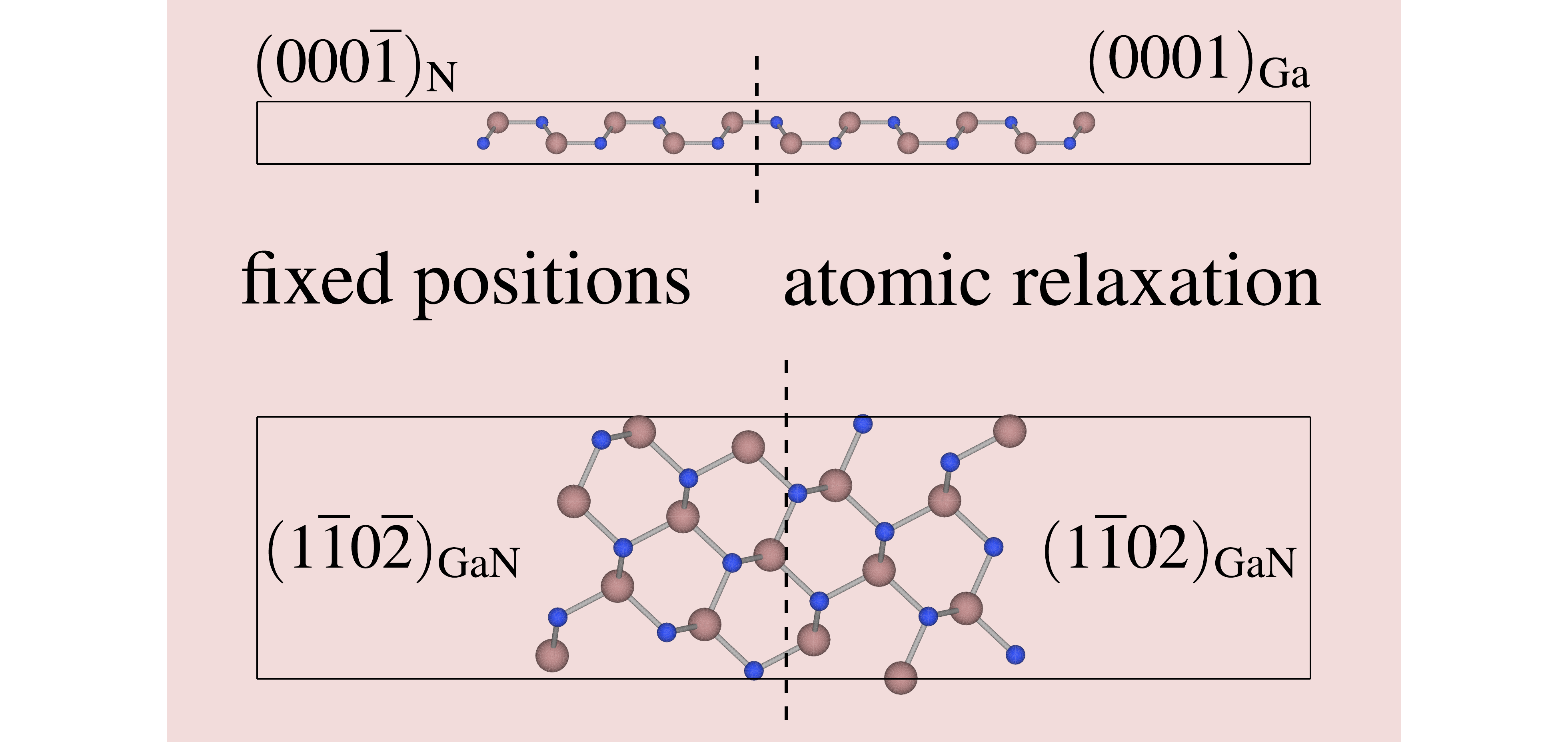}}
\caption{(color online) Structures to calculate particular surface-energy combinations according to Fig.~\ref{fig:workflow}. Top: Wedge structure with one $(000\overline{1})_\text{N}$ facet and two $(1\overline{1}02)_\text{GaN}$ facets. Middle: Slab along the $[1\overline{1}02]$ direction. Bottom: Additional slabs, allowing for atomic relaxations on the $(0001)_\text{Ga}$ and  $(1\overline{1}02)_\text{GaN}$ surfaces. Vertical dashed lines indicate the boundaries between the fixed part and the relaxed part of the slab. Large (small) spheres represent Ga (N) atoms. \label{fig:exam}}
\end{figure}

Having achieved this goal, as illustrated above, we need to show now that such a surface-energy combination $\Delta \gamma_i$ can be indeed calculated for each semipolar plane. Table~\ref{tab:sigma} lists all involved surface-energy combinations calculated from wedges only (I) or in combination with slabs (II and III). Note that $\sigma$ is the surface energy per surface cell and is related to $\gamma$ by $\sigma=\gamma A$. The surface-energy combinations I and II were calculated for unrelaxed (superscript "un") surfaces with $H$ passivation, and the combinations III are, consequently, derived involving surface relaxations. The used wedge structures can be found in the SM \cite{supp}. The overall scheme of our calculations, as sketched in Fig.~\ref{fig:workflow}, involves five steps. (1) The most stable surface terminations of one semipolar plane and one polar plane are adopted to build the 1D wedge structures (as shown in the upper panel of Fig.~\ref{fig:exam}, see also the SM \cite{supp}). This allows us to determine the sum of surface energies of a semipolar surface and a polar surface. (2) Using a 2D slab, the sum of two surface energies is obtained for different slab orientations. (3) Combining the sum of surface energies obtained from the wedge (I) with the sum of surface energies obtained from the slab, combination II is derived. (4) Additional calculations are carried out with the slab approach. However, this time, one of the two surfaces of a slab is relaxed while the opposite surface is kept fixed and $H$ passivated as in previous cases of slabs and wedges. That way, the sum of surface energies of a relaxed surface and an unrelaxed surface is obtained. (5) Finally, these surface-energy sums are added to combinations II to achieve combinations of type III. Figure~\ref{fig:exam} illustrates a particular example for determining the surface-energy combination $\sigma_{1\overline{1}02}-2\sigma_{0001}$ for the case of the $(1\overline{1}02)_\text{GaN}$ surface and the $(0001)_\text{Ga}$  surface. Results for other crystal planes, as well as different surface configurations are calculated accordingly. 

\begin{table}
\caption{Surface-energy combinations as obtained from the wedge calculations only (I) or in combination with slab calculations (II, III). (a), (b), and (c) refer to different wedge structures (see the SM \cite{supp}). Combinations in every second row refer to wedge structures with interchanged Ga and N, required for the construction of the lower half of the ECS.}
\centering
\begin{tabular*}{\linewidth}{@{\extracolsep{\fill}} c  c  c  c  c  c  c }
\hline \hline
Wedge & \multicolumn{2}{c}{I} & \multicolumn{2}{c}{II} & \multicolumn{2}{c}{III} \\
\hline
(a) & \multicolumn{2}{c}{$\sigma_{11\overline{2}2}^\text{un}+2\sigma_{000\overline{1}}^\text{un}$} & \multicolumn{2}{c}{$\sigma_{11\overline{22}}^\text{un}-2\sigma_{000\overline{1}}^\text{un}$} & \multicolumn{2}{c}{$\sigma_{11\overline{2}2}-2\sigma_{0001}$} \\
& \multicolumn{2}{c}{$\sigma_{11\overline{22}}^\text{un}+2\sigma_{0001}^\text{un}$} & \multicolumn{2}{c}{$\sigma_{11\overline{2}2}^\text{un}-2\sigma_{0001}^\text{un}$} & \multicolumn{2}{c}{$\sigma_{11\overline{22}}-2\sigma_{000\overline{1}}$} \\
(b) & \multicolumn{2}{c}{$\sigma_{1\overline{1}01}^\text{un}+\sigma_{000\overline{1}}^\text{un}$} & \multicolumn{2}{c}{$\sigma_{1\overline{1}0\overline{1}}^\text{un}-\sigma_{000\overline{1}}^\text{un}$} & \multicolumn{2}{c}{$\sigma_{1\overline{1}01}-\sigma_{0001}$} \\
& \multicolumn{2}{c}{$\sigma_{1\overline{1}0\overline{1}}^\text{un}+\sigma_{0001}^\text{un}$} & \multicolumn{2}{c}{$\sigma_{1\overline{1}01}^\text{un}-\sigma_{0001}^\text{un}$} & \multicolumn{2}{c}{$\sigma_{1\overline{1}0\overline{1}}-\sigma_{000\overline{1}}$} \\
(c) & \multicolumn{2}{c}{$\sigma_{1\overline{1}02}^\text{un}+2\sigma_{000\overline{1}}^\text{un}$} & \multicolumn{2}{c}{$\sigma_{1\overline{1}0\overline{2}}^\text{un}-2\sigma_{000\overline{1}}^\text{un}$} & \multicolumn{2}{c}{$\sigma_{1\overline{1}02}-2\sigma_{0001}$} \\
& \multicolumn{2}{c}{$\sigma_{1\overline{1}0\overline{2}}^\text{un}+2\sigma_{0001}^\text{un}$} & \multicolumn{2}{c}{$\sigma_{1\overline{1}02}^\text{un}-2\sigma_{0001}^\text{un}$} & \multicolumn{2}{c}{$\sigma_{1\overline{1}0\overline{2}}-2\sigma_{000\overline{1}}$} \\
\hline \hline
\end{tabular*}
\label{tab:sigma}
\end{table}

\begin{figure}
\includegraphics[width=\linewidth]{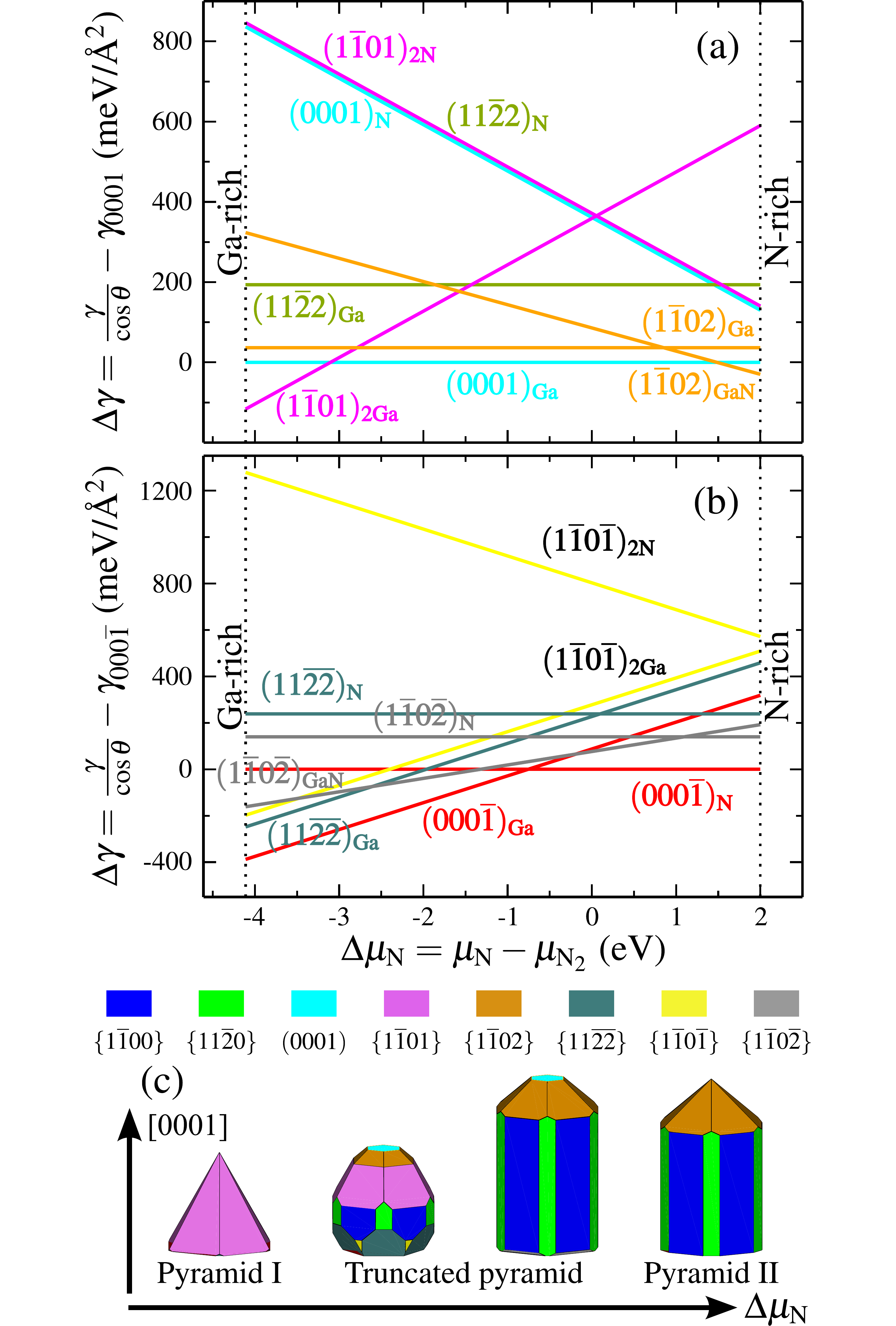}
\caption{\label{fig:crystal} (color online) Surface energies and crystal shapes of wurtzite GaN. (a) Relative energies $\Delta \gamma$ versus chemical potential along the [0001] direction. (b) The same but for the $[000\overline{1}]$ direction. (c) GaN crystals under thermodynamic equilibrium conditions. The shape varies continuously from Ga-rich conditions (left) to N-rich conditions (right).  The labels---pyramid I, truncated pyramid, and pyramid II---refer to the top shape.}
\end{figure}

These quantities are plotted in Fig.~\ref{fig:crystal}. Typically, the chemical potential of nitrogen $\mu_\text{N}$ can vary from Ga-rich to N-rich conditions, $E_\text{GaN}-E_\text{Ga} \leq \mu_\text{N} \leq E_{\text{N}_2}$, where $E_\text{GaN}$ and $E_\text{Ga}$ represent the total energies of wurtzite GaN and bulk Ga, respectively, and $E_{\text{N}_2}$ is the total energy (per atom) of the $\text{N}_2$ molecule. In the figure, this range is extended by 2~eV on both sides to mimic experimental temperature and pressure conditions. We find that the $(0001)_\text{N}$ termination has higher surface energy than the $(0001)_\text{Ga}$ termination over the whole considered range of $\Delta \mu_\text{N}$. $\Delta \gamma$ of the $(1\overline{1}01)_\text{2Ga}$ surface becomes negative when $\Delta \mu_\text{N}$ is low. On the other hand, $\Delta \gamma$ of the $(1\overline{1}02)_\text{GaN}$ surface becomes negative when $\Delta \mu_\text{N}$ is high. Therefore, $(1\overline{1}01)_\text{2Ga}$ and $(1\overline{1}02)_\text{GaN}$ surfaces dominate the Wulff construction in these two extreme cases. For an intermediate range of $\Delta \mu_\text{N}$, the $(0001)_\text{Ga}$ surface has the lowest energy; thus, this surface remains at the top of the crystal. On the bottom side, as shown in Fig.~\ref{fig:crystal}(b), the situation is different: The $(000\overline{1})_\text{Ga}$ surface is more stable than the semipolar surfaces, and for higher $\Delta \mu_\text{N}$ $(000\overline{1})_\text{N}$ becomes more favorable. This means that the crystal is terminated either by the $(000\overline{1})_\text{Ga}$ or the $(000\overline{1})_\text{N}$ facet under various $\Delta \mu_\text{N}$ conditions.

Figure~\ref{fig:crystal}(c) shows the 3D Wulff crystals constructed from the above results. We summarize the major features. (i) The GaN crystal, in general, exhibits a rodlike shape along [0001] under various conditions. (ii) The shapes on the top and bottom side are changing according to the chemical potential. Under extremely Ga-rich conditions, the crystal forms a complete pyramid consisting of $\lbrace1\overline{1}01\rbrace$ planes at the top. When $\Delta\mu_\text{N}$ increases to a higher value, the crystal adopts the shape of a truncated pyramid, and eventually turns into another pyramid formed by $\lbrace1\overline{1}02\rbrace$ facets. At the bottom side, the flat $(000\overline{1})$ surface turns into a polyhedral shape consisting of $(000\overline{1})$, $\lbrace11\overline{22}\rbrace$, and $\lbrace1\overline{1}0\overline{1}\rbrace$ planes, and is further formed by $\lbrace1\overline{1}0\overline{2}\rbrace$ and $(000\overline{1})$ planes only. Finally, the flat $(000\overline{1})$ plane appears again. (iii) The side wall consists of both nonpolar facets, namely the \textit{m} plane and the \textit{a} plane. These findings complement those of Lymperakis and Neugebauer \cite{Lymperakis2009a}, having shown that the coexistence of these facets facilitates two diffusion channels for Ga atoms. When Ga atoms arrive at the \textit{m} plane (indicated by the blue color), lateral diffusion is favorable; for Ga atoms at the \textit{a} planes (green color), vertical diffusion along the \textit{c} axis takes place. Ga atoms then accumulate at the top of the crystal and axial growth continues. The appearance of two nonpolar planes on the side walls also agrees with recent experiments \cite{Urban2013,Brandt2014} that nanocolumns and NWs indeed do not exhibit atomically sharp corners.

Finally, let us recall the variety of experimentally achieved crystal morphologies. Selective-area MOCVD growth \cite{Du2005} exhibited convex $\lbrace1\overline{1}01\rbrace$ and concave $\lbrace11\overline{2}2\rbrace$ surfaces. In contrast, Jindal \cite{Jindal2009} observed in MOCVD growth a complete hexagonal pyramid on the $(0001)$ plane as its equilibrium shape, and truncated hexagonal pyramids out of equilibrium, while the crystal grown on the $(11\overline{2}0)$ and $(1\overline{1}00)$ planes showed $\lbrace1\overline{1}01\rbrace$ facets along the $[0001]$ direction and a $(000\overline{1})$ facet on the opposite side. In hydride vapor phase epitaxy \cite{Bryant2013}, depending on the temperature and pressure, the truncation of the pyramidal shape was confirmed to be continuously varying along the [0001] direction. More recently, semipolar $\lbrace1\overline{1}02\rbrace$ facets on top of GaN nanocolumns were reported from selective-area MBE growth \cite{Urban2013}. Considering our theoretical results, these observations are not controversial. In fact, most of the observed crystal morphologies are consistent with our computed ECSs. Particularly, the crystal shape under N-rich conditions is fully consistent with the different morphologies of GaN NWs \cite{Fernandez2012,Chen2010,Mandl2013}, where pyramid or truncated pyramid shapes are observed in the case of Ga-polar NWs, and the flat $(000\overline{1})$ facet dominates the top in the case of N-polar NWs. At the same time, the aspect ratio of experimental NWs can be much larger than that seen in these ECSs. This implies that kinetic effects also play an important role in NW growth.

Summarizing, we have demonstrated a generalization of Wulff construction to determine equilibrium crystal shapes for wurtzite crystals.  Although the individual surface energies for semipolar and polar surfaces are not accessible, the ECS can still be obtained. For each semipolar plane, the relative energy with respect to its neighboring polar plane can be unambiguously computed as a function of chemical potential. This energy difference, corresponding to the crystal radius along the polar axis, is the important quantity that governs the crystal shape. We have exemplified our approach with wurzite GaN. Taking into account several bulk-truncated surfaces, ECSs have been constructed. These crystals exhibit a rodlike shape along the polar \textit{c} axis, with top and bottom geometries depending on the chemical potential, while the side walls are formed by both types of nonpolar surfaces. Our results can well explain the experimentally observed NW shapes. They also open a perspective to gaining insight into morphologies of the entire class of polar materials, concerning point groups of 6, 6$mm$, 4, 4$mm$, 3, 3$mm$, 2, and 2$mm$, where such polar axes exist.

Input and output files of our calculations can be downloaded from the NoMaD Repository by following the link in Ref. \cite{nomad}.

\begin{acknowledgements}
We gratefully acknowledge valuable discussions with S. Fern\'{a}ndez-Garrido and O. Brandt, and we thank S. Erwin for critical reading of the manuscript.
\end{acknowledgements}


\begin{thebibliography}{26}
\bibitem{shapes} The assumption that all surface energies are lowered by surface reconstruction by the same amount would obviously not have any impact on the shape. The fist results on the reconstruction of the (0001) and ($000\overline{1}$) surfaces show a shrinkage of the pyramid tip, while the overall shape is retained. Considering the effect of reconstruction on the ($1\overline {1}02$) surface basically recovers the original ECS.
\end{thebibliography}
\end{document}